\documentclass[graybox]{svmult}

\usepackage{mathptmx}       % selects Times Roman as basic font
\usepackage{helvet}         % selects Helvetica as sans-serif font
\usepackage{courier}        % selects Courier as typewriter font
\usepackage{type1cm}        % activate if the above 3 fonts are
\usepackage{amsmath,latexsym,amssymb}
\usepackage{graphicx,qtree,bussproofs}

\usepackage{makeidx}         % allows index generation
\usepackage{multicol}        % used for the two-column index
\usepackage[bottom]{footmisc}% places footnotes at page bottom

\makeindex             % used for the subject index
\begin{document}

\title*{A Dynamic-Epistemic Logic for Mobile Structured Agents}
\author{Anya Yermakova and Alexandru Baltag}
\institute{Anya Yermakova \at Mathematics and Computer Science, Oxford University \email{ayermakova@gmail.com}}
\maketitle

\abstract{Multi-agent systems have been studied in various contexts of both application and theory. We take Dynamic Epistemic Logic (DEL), one of the formalisms designed to reason about such systems, as the foundation of the language we will build.
\newline\indent
BioAmbient calculus is an extension of $\pi$-calculus, developed largely for applications to biomolecular systems. It deals with ambients and their ability to communicate and to execute concurrent processes while moving.
\newline\indent
In this paper we combine the formalism of Dynamic Epistemic Logic together with the formalism of BioAmbient Calculus in order to reason about knowledge maintained and gained upon process transitions. The motivation lies in developing a language that captures locally available information through assignment of knowledge, with potential application to biological systems as well as social, virtual, and others.
\newline\indent
We replace the ambients of BioAmbient Calculus with agents, to which we attribute knowledge, and explore the parallels of this treatment. The resulting logic describes the information flow governing mobile structured agents, organized hierarchically, whose architecture (and local information) may change due to actions such as \textit{communication}, \textit{merging} (of two agents), \textit{entering} (of an agent into the inner structure of another agent) and \textit{exiting} (of an agent from the structure of another).  We show how the main axioms of DEL must be altered to accommodate the informational effects of the agents' dynamic architecture.}

\section{Introduction}
\label{Introduction}

We develop a formalism $\mathcal{PADEL}$ suited for talking about various multi-agent systems. In particular, we discuss previous and potential applications to systems of molecular biology, though the language is not limited to this. We develop the notion of an \textit{agent}, which can refer to an entire system or a subsystem thereof, all seen as \textit{informational} (and \textit{information-acquiring}) systems. Information locally available to a given system is treated as \textit{knowledge} and the flow and exchange of information between systems as \textit{dynamics of knowledge} in a multi-agent setting. For all of the above, we rely on a formalism derived from Dynamic Epistemic Logic and BioAmbient Calculus.
\par
We assume the following things about the architecture of these agents: First, the number of agents (and thus subagents) is always finite. Second, they are nested in a dynamic tree structure (with no loops). 
\par
In addition to typical communication actions, such as sending and receiving information or public announcements, we consider three specific actions which involve mobility: entering, exiting, and merging. The formalisation and the specific rules for the latter are inspired largely by Luca Cardelli's developments in BioAmbient Calculus\footnote{See \cite{Bioware} and \cite{MobileAmbients}}, which aims to formalize information flow in systems of molecular biology.

\section{The formalism and motivation of $\mathcal{PADEL}$}
\label{syntax and semantics}
At a given state, an agent is to be defined by an assignment of concurrent processes, and in a given process there can occur agents, capabilities, or other non-agent, non-capability processes. 

\subsection{Basic Definitions}
Let $\mathcal{A}$ be a finite set of agents and $\mathcal{A}c$ a finite set of atomic actions.\\

An agent $A \in \mathcal{A}$ occurs in a process $P$, or $A\sqsubseteq P$, iff $\begin{cases}  A \sqsubseteq A \\ A\sqsubseteq P \Rightarrow A\sqsubseteq P \mid Q \\  A\sqsubseteq P \Rightarrow A\sqsubseteq Q\mid P \\  A\sqsubseteq P \Rightarrow A\sqsubseteq a.P  \end{cases}$

where $P \mid Q$ denotes two processes running in parallel and $a.P$ denotes an action capability $a$, which, if executed, will initiate a process $P$. 
\par
We define $\sqsubseteq^{+}$ as the transitive closure of $\sqsubseteq$: 
\par
$A\sqsubseteq^{+}P \ \Leftrightarrow \ \exists $ a chain $ P_{0}, P_{1}, ..., P_{n} $ of processes 
s.t. $n>0$, 
$A=P_{0}, \ P=P_{n}$, and $P_{i-1}\sqsubseteq P_{i}$, for all $i\leq n$.\\
\par
% \newline
An agent $A\in \mathcal{A}$ is a \textit{subagent} of $P$, or $A<P$, iff $\begin{cases}  A < A \mid P \\ A < P \mid A \\ A< P \Rightarrow A< P \mid Q \\  A < P \Rightarrow A<Q\mid P  \end{cases}$

\runinhead{Definition}
A \textit{state} $s$ is an assignment of Processes to Agents, $s: Agents \rightarrow Processes$, such that for every two distinct agents $A,B \in \mathcal{A}$ and for any agent $C \in \mathcal{A}$: 
\begin{equation}
C \sqsubseteq s(A), C \sqsubseteq s(B) \Rightarrow A=B \mbox{ and } A \not\sqsubseteq^{+} s(A) \label{pre*} 
\end{equation}

%\begin{equation}
%C \sqsubseteq s(A), C \sqsubseteq s(B) \Rightarrow A=B, \ and \label{pre*}
%\end{equation}
%\begin{equation}
%A \not\sqsubseteq^{*} s(A) \label{refl}
%\end{equation}

That is, agent $C$ cannot simultaneously occur in a process assigned to two different agents and an agent cannot occur in a process assigned to itself.

%\newline
For a given state $s$ and two agents $A, B$, we define $A<_s B \stackrel{def}{=} A < s(B)$. We read $A<_s B$ as ``$A$ is a subagent of (agent) $B$ in state $s$.''

%\runinhead{Definition}
%We let $s(B)$ represent all processes, running concurrently, that are assigned to agent $B$ at state $s$. For a given state $s$,  $A<_{s}B \stackrel{def}{=} A<s(B)$. 

\runinhead{Consequences}

\begin{equation}
A<_{s}B \Rightarrow A \sqsubseteq s(B). \label{p22a}
\end{equation}

From \eqref{pre*} and \eqref{p22a}, it follows also that: 
\begin{equation}
C <_{s} A, C <_{s} B \Rightarrow A=B \label{*}
\end{equation}

In other words, assignments $s(A)$ and $s(B)$ for different agents $A\not=B$ must contain no agents in common. 

\runinhead{Definition}
We define $<_{s}^{+}$ as the transitive closure of $<_{s}$, and call it the $iterative$ subagent relation at state $s$, while referring to $<_{s}$ as the $one-step$ subagent relation.  
\par
$A<_{s}^{+}B \ \Leftrightarrow \ \exists $ a finite chain $ A_{0}, A_{1}, ..., A_{n} $ \\ 
s.t. $n>0$, 
$A=A_{0}, \ B=A_{n}$, and $A_{i-1}<_{s}A_{i}$, for all $i\leq n$.

Consequence \ref{*} in turn disallows loops in the tree of agents:
\runinhead{Proposition [Tree Property]}
For $A \neq B$: 
\begin{equation}
C<^{+}_{s}A, \ C<^{+}_{s}B \ \Rightarrow \ A<^{+}_{s}B \vee B<^{+}_{s}A . \label{treelong} 
\end{equation}

\begin{proof}
\smartqed
We prove this by induction on the length of the chain. By the hypothesis, there must exist two chains, where $A\neq B$:\\
$ C=X_{1}<_{s}X_{2}<_{s}...X_{i-1}<_{s}X_{i}<_{s}X_{i+1}<_{s}...<_{s}X_{n}=A$ and \\
$ C=Y_{1}<_{s}Y_{2}<_{s}...Y_{i-1}<_{s}Y_{i}<_{s}Y_{i+1}<_{s}...<_{s}Y_{m}=B$.\\
Without loss of generality, suppose $n\leq m$ (the case for $m<n$ is similar).\\
Then, by \eqref{*}, $X_{2}=Y_{2}$, and again by \eqref{*}, $X_{3}=Y_{3}$, and so on until $X_{n}=Y_{n}, \ \Rightarrow \ A=Y_{n}$. \\
Now, if $n=m$, then $A=Y_{m}=B$, contradicting the fact that $A$ and $B$ were assumed to be distinct.\\
If $n<m$, then $A=Y_{n}<_{s}Y_{n+1}<_{s}...<_{s}Y_{m}=B$ and we have shown that $A<_{s}^{+}B$. 
\qed
\end{proof}

Given a finite set $\mathcal{A}$ of agents, denoted by $A,B,C,A_{1},..., A_{n}$, and given a finite set $\mathcal{A}c$ of atomic actions, denoted by $a,a_{i}, \overline{a}$, we combine the syntax of BioAmbient Process Algebra and DEL, adding only the atomic sentence $A<^{+}B$, and define the sentences of propositional logic together with the $one-step$ subagent relation. $\varphi, \psi$ are formulae and  $p$ are propositional sentences in the language:  

\begin{table}
\caption{Syntax and Definitions}
\label{syntaxtable}  
\begin{svgraybox}

%\begin{center}
\begin{tabular}{l @{}} 
Assume $A,B,C$ are distinct agents. Then: \\
\\
\hline
\\
P ::$=\ \textbf{0}\ \mid \ A \ \mid \ (P\mid P)\ \mid \  \Sigma_{i} a_{i}.P_{i} $\\
$\varphi::= \ A<^{+}B \mid \neg\varphi \mid \varphi \wedge \varphi \mid K_{A}\varphi \mid DK_{A_{1},..., A_{n}}\varphi \mid [\alpha]\varphi$ \\
$\alpha \in \{(a_{A}, \overline{a}_{B})\}$ \\
$a, a_{i}, \overline{a} \in \{\varphi?, \ \varphi!, \ enter, \ accept, \ exit, \ expel, \ merge+ , \ merge-\}$ \\
\hline
\end{tabular}
%\end{svgraybox}
%
%\end{table}
%
%
%\begin{table}
%\caption{Definitions}
%\label{definitionstable}
%\begin{svgraybox}

\begin{tabular}{r c l @{}}
\\
$\varphi \vee \psi$ & $:\stackrel{def}{=} $ & $ \neg(\neg\varphi \wedge \neg\psi)$\\
$\varphi \Rightarrow \psi$ & $:\stackrel{def}{=} $ & $ \neg (\varphi \vee \psi)$\\
$\varphi \Leftrightarrow \psi$ & $:\stackrel{def}{=} $ & $ (\varphi \Rightarrow \psi) \wedge (\psi \Rightarrow \varphi)$\\
$ A<C $ & $ :\stackrel{def}{=} $ & $ A<^{+}C \ \wedge \ \bigwedge_{B \in \mathcal{A}} \neg (A<^{+}B \ \wedge \ \ B<^{+}C)$ \\
$<\alpha>\varphi $ & $:\stackrel{def}{=} $ & $ \neg[\alpha]\neg\varphi$\\
$\top$ & $:\stackrel{def}{=} $ & $p \vee \neg p$, for some fixed $p$ \\
$\bot$ & $:\stackrel{def}{=} $ & $p \wedge \neg p$, for some fixed $p$ \\
$$\\
\hline
\end{tabular}
\end{svgraybox}
\end{table}

\subsection{Actions}
\label{Actions}
The set $\mathcal{A}c$ of atomic actions is finite. Similar to the notions of $executability$, or $precondition$, in DEL, agent $A$ must have the capability $a.P$ included in the processes assigned to it at the initial state in order for $a_{A}$ to take place (agent $A$ executing action $a$).
\par
The capabilities each agent is assigned at a given state are expressed as a nondeterministic sum of atomic actions 
$\sum\limits_{i} a_{i}.P_{i}$, each of which is attached to the process that would initiate as a result of $A$ performing a given atomic action.

\subsubsection{State Transitions}
\label{transitions}
Following the Bioambient improvement on Ambient Calculus, we only allow suitable action \textit{pairs} to induce state transitions. A cell has to accept a virus that is trying to enter, just like an announcement must be heard in order for it to affect an agent's knowledge.
\par
We define actions $\alpha$ as dual pairs of atomic actions, which form a finite set $i\mathcal{A}c$:
\begin{center}
$\alpha=(a,\overline{a}) \in \mathcal{A}c$ x $\mathcal{A}c = i\mathcal{A}c$. 
\end{center}

%Type I - actions of mere communication (such as $send$ and $receive$). \\
%Type II - actions where $a=enter , \ \overline{a} = accept $, and any other pairs $(a_{i}, a_{j}) \in i\mathcal{A}c$ that behave the same way. \\
%Type III - actions where $a=exit , \ \overline{a} = expel$, and any other $(a_{i}, a_{j}) \in i\mathcal{A}c$ that behave the same way. \\
%Type IV - actions where $a=merge + , \ \overline{a} = merge - $, and any other $(a_{i}, a_{j}) \in i\mathcal{A}c$ that behave the same way. \\

% Note also that the composition of the set $i\mathcal{A}c^{+}$ changes at every state assignment, depending on current capabilities assigned to agents $A,B$ at a given state. Assigning particular agents to our actions makes them deterministic and avoids the problem of deciding which agents will perform an action if several have the same capability. However, we do not disregard the fact that different agents can have the same capability, and that many different $\alpha$'s can be chosen from $i\mathcal{A}c^{+}$ to execute the same type of action. This is different from the ambient formalism, where if a capability exists in several ambients of the same name, one of them is chosen nondeterministically to carry out the action process. 
%\par
We use $B:_{s}\alpha$ to denote agent $B$'s participation in action $\alpha$ at state $s$. For $\alpha=(a_{A},\overline{a_{C}})$: 
% For $\alpha = (a_{A}, \overline{a}_{C})$: \\
\par
$B:_{s}\alpha \ \stackrel{def}{=} \ A\leq^{+}_{s}B \ \vee \ C\leq^{+}_{s}B$ and $ \exists s'$ s.t. $s \stackrel{\alpha}{\rightarrow}s'$ \\
\par
We define four types of actions, of which three involve a $one-step$ superagent $E$ whose state assignment is crucial to the executability of the action (see Figure 1). 
\begin{center}
% \\
For any agents $A,C,E$ that are distinct:
\\
\begin{tabular}{l @{}}
$\alpha_{I}= (\varphi?_{A}, \varphi!_{C})$, \\
$\alpha_{II}= (enter_{A}, accept_{C},E)$, \\
$\alpha_{III}= (exit_{A}, expel_{C},E)$, \\
$\alpha_{IV}= (merge+_{A}, merge-_{C},E)$ \\
\end{tabular}
\end{center}

\begin{figure}
\sidecaption
% For example, with the graphicx style use
\includegraphics[scale=.29]{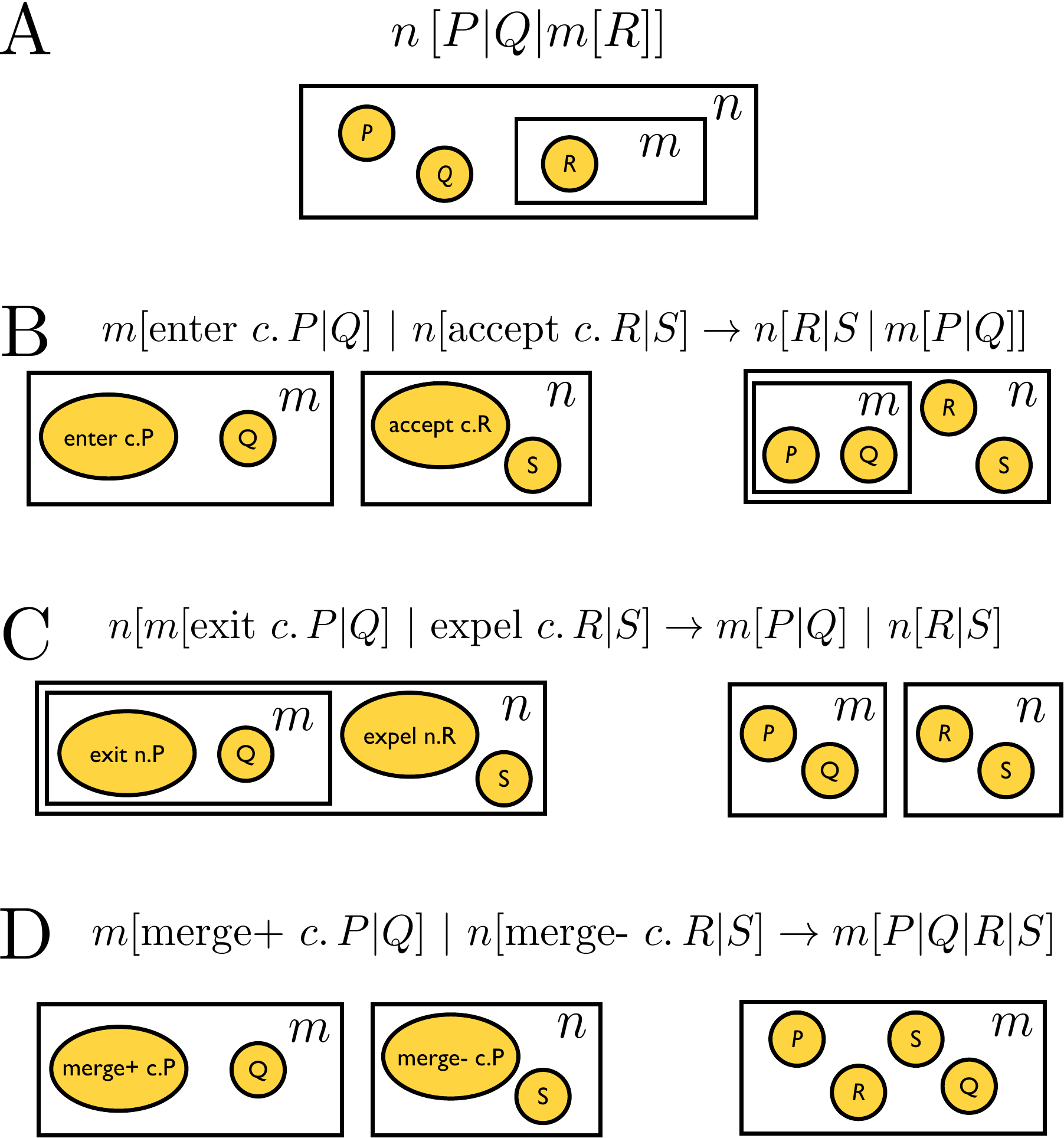}
%
% If no graphics program available, insert a blank space i.e. use
%\picplace{5cm}{2cm} % Give the correct figure height and width in cm
%
\caption{Motivated by work of Luca Cardelli (see \cite{carvirus}-\cite{MobileAmbients}), this figure depicts the application of this language to molecular biology. B, C, and D show the change in structure of processes and subprocesses as a result of acting on dual capabilities (Types II, III, IV, respectively), separated by no more than two ''membranes.'' In $\mathcal{PADEL}$, we can think of each "membrane" with all its contents as a unique agent.}
%\label{mobilityfigure}       % Give a unique label
\end{figure}

We now define the state transitions for the four different types of actions.

\runinhead{Type I} For $\alpha_{I}=(a_{A}, \overline{a}_{C})$, where $(a,\overline{a}) \ = \ (\varphi?, \varphi!)$:
\par
$s\stackrel{\alpha_{I}}{\rightarrow}s'$
iff \\
$\exists \ a, a_{i}, P, P_{i}, Q \mbox{ such that }s(A) = \sum_{i}a_{i}.P_{i}+a.P \mid Q$,\\
$\exists \ \overline{a}, c_{j}, R, R_{j}, S \mbox{ such that }s(C) = \sum_{j}c_{j}.R_{j}+\overline{a}.R \mid S$,\\
$s'(A) = P \mid Q, \ 
s'(C) = R \mid S, \ 
s'(X) = s(X), \mbox{ for all }X \neq A,C$. \par
This is the only type of action that does not change the structure of the tree of agents.

\runinhead{Type II} For $\alpha_{II} = (a_{A}, \overline{a}_{C},E)$ where $a = enter , \ \overline{a} = accept $:  
\par
s$\stackrel{\alpha_{II}}{\rightarrow}$s$'$
iff \\
$\exists \ a, a_{i}, P, P_{i}, Q \mbox{ such that }s(A) = \sum_{i}a_{i}.P_{i}+a.P \mid Q$,\\
$\exists \ \overline{a}, c_{j}, R, R_{j}, S \mbox{ such that }s(C) = \sum_{j}c_{i}.R_{j}+\overline{a}.R \mid S$,\\
$\exists \ \Gamma, \ a \ process, \mbox{ such that }s(E)=A \mid C \mid \Gamma$, and \\
$s'(A) = P \mid Q$,\
$s'(C) = A \mid R \mid S$,\
$s'(E) = C \mid \Gamma$,\
$s'(X) = s(X), \mbox{ for all }X \neq A,C,E$.
\par 
After $\alpha_{II}$ state transition, agent $C$ is assigned a new agent, while agent $E$ -- the initial superagent of both $C$ and $A$ -- is stripped of the $one-step$ subagent $A$:
\parbox[t]{2.4in}{ ~\Tree 
   [.E A C  ] }
   $\stackrel{\alpha_{II}}{\Longrightarrow}$ 
\parbox[t]{2in}{ \Tree [.E [.C {A} ]]
  }
  
\runinhead{Type III} For $\alpha_{III} = (a_{A}, \overline{a}_{C},E)$ where $a = exit , \ \overline{a} = expel $: 
\par
s$\stackrel{\alpha_{III}}{\rightarrow}$s$'$
iff \\
$\exists \ a, a_{i}, P, P_{i}, Q \mbox{  such that  }s(A) = \sum_{i}a_{i}.P_{i}+a.P \mid Q$,\\
$\exists \ \overline{a}, c_{j}, R, R_{j}, S \mbox{ such that }s(C) = \sum_{j}c_{i}.R_{j}+\overline{a}.R \mid A \mid S$,\\
$\exists \ \Gamma, \ a \ process, \mbox{ such that }s(E)=C \mid \Gamma$, and \\
$s'(A) = P \mid Q, \ 
s'(C) = R \mid S, \ 
s'(E) = C \mid A \mid \Gamma,\ 
s'(X) = s(X), \mbox{ for all }X \neq A,C,E$.
\par
After $\alpha_{III}$ this state transition, the effect is exactly opposite to that of transitions by actions of Type II: \\ 

\parbox[t]{2in}{ \Tree [.E [.C {A} ]]
  }
 $\stackrel{\alpha_{III}}{\Longrightarrow}$ 
\parbox[t]{2.4in}{ ~\Tree 
   [.E A C  ] }

\runinhead{Type IV} $\alpha_{IV}$, defined as $(a_{A}, \overline{a}_{C},E)$ where $a = merge+ , \ \overline{a} = merge- $:  
\par
s$\stackrel{\alpha_{IV}}{\rightarrow}$s$'$
iff \\
$\exists \ a, a_{i}, P, P_{i}, Q \mbox{ such that }s(A) = \sum_{i}a_{i}.P_{i}+a.P \mid Q$,\\
$\exists \ \overline{a}, c_{j}, R, R_{j}, S \mbox{ such that }s(C) = \sum_{j}c_{i}.R_{j}+\overline{a}.R \mid S$,\\
$\exists \ \Gamma, \ a \ process, \mbox{ such that }s(E)=A \mid C \mid \Gamma$, and \\
$s'(A) = P \mid Q \mid R \mid S,\  
s'(C) =  \textbf{0},\ 
s'(E) = A \mid \Gamma,\ 
s'(X) = s(X), \mbox{ for all }X \neq A,C,E$.
\par
\parbox[t]{2.4in}{ ~\Tree 
   [.E A C  ] }
   $\stackrel{\alpha_{IV}}{\Longrightarrow}$ 
\parbox[t]{2in}{ \Tree [.E  A ] }
\par
The following validities follow immediately from the definitions, where, as in DEL, \\ 
$<\alpha_{i}>\top$ denotes executability of $\alpha_{i}$, and $[\alpha_{i}]\varphi$ denotes a statement $\varphi$ that holds true after action $\alpha_{i}$: 
\begin{table}
\caption{Consequences of Action Definitions}
\begin{svgraybox}
\begin{tabular}{l c @{}}
Assume $A,C,E$ are distinct agents. Then:\\
\\
\hline
$<\alpha_{II}>\top \ \Rightarrow \ A<E \wedge C<E$ & $$\\
$<\alpha_{III}>\top \ \Rightarrow \ A<C \wedge C<E$ & $$\\
$<\alpha_{IV}>\top \ \Rightarrow \ A<E \wedge C<E$ & $$\\
$[\alpha_{II}]A<C $ & $$\\
$[\alpha_{III}]A<E $  & $$\\
$[\alpha_{IV}]\neg C<E $  & $$\\
\hline
\end{tabular}
\end{svgraybox}
\end{table}
\newpage

\subsection{Indistinguishability Relations on States and Actions}
As in Epistemic Logic (EL), we define indistinguishability of two states for a particular agent $A$, denoted by $\sim_{A}$, in order to reason about knowledge. 
\runinhead{Definition}[Indistinguishability of states] 
$s\sim_{A}s' $ iff $ s(X)=s'(X),$ for all $X \leq^{+}_{s} A$. 
\par
That is, two states are equivalent for agent $A$ if and only if they are indistinguishabe for $A$ and all of its subagents, as assigned at state $s$. 
%The next statement follows from this Definition:
%\begin{equation}
%s\sim_{C}s', \ A<^{+}_{s} \ \Rightarrow \ s\sim_{A}s' \label{zhuk-s}
%\end{equation}\par
Two states can be indistinguishable for a group of agents $B_{1},...,B_{n}$ if none of them can distinguish between these states: 
\runinhead{Definition}[Indistinguishability of states for a group of agents] $s\sim_{B_{1},...,B_{n}}s' $ iff $ s(X)=s'(X),$ for all $X \leq^{+}_{s}B_{i}$, for all $1\leq i \leq n$. 
\par
We define indistinguishability of actions: 
\runinhead{Definition}[Equivalence of actions] Here $s$ represents any state in the history.
\par
$\alpha \stackrel{A}{\sim}_{s} \alpha' \Leftrightarrow 
\begin{cases} either \ A:_{s}\alpha$ and $\alpha=\alpha' \\
or \ \neg A:_{s}\alpha \ and \ \neg A:_{s}\alpha'\  
\end{cases}$ \\
\par
If $A$ is a participant in $\alpha$, then it would certainly be able to differentiate between taking part in two different actions $\alpha$ and $\alpha'$, unless they were actually the same.
On the other hand, if $A$ does not participate in either $\alpha$ ot $\alpha'$, then both actions appear equivalent to $A$. This implies that $A$ is a subagent of both agents executing $\alpha$.

\section{Semantics}
\label{Semantics}
We will evaluate logical formulas on \textit{histories}, which are sequences of states and actions (representing possible histories of a system). However, in order to define the semantics for epistemic and dynamic modalities, we need to define appropriate (epistemic) indistinguishability relations and (dynamic) transition relations on histories, by lifting to histories the corresponding state relations.
%As in Dynamic Epistemic Logic, ``$s\models \varphi$'' represents a statement $\varphi$ valid at state $s$, and ``$s\models \varphi$ iff $\psi$'' provides the semantical description $\psi$ of what holds true at $s$. In particular:
%\runinhead{Proposition}
%$s\models \ A<^{+}B $ iff $ A<^{+}_{s}B$.

\subsection{Relations on Histories}
To ensure that our knowledge is accumulative, as in DEL, we must expand the language to include Perfect Recall and extend equivalence relations to state transitions and to previous states. For this we define histories and develop axioms based on histories rather than states.
\par
We define a \textit{history} $h$ as a sequence of alternating states and actions: 
\begin{eqnarray}
h = (s_{0}, \alpha_{0}, s_{1}, \alpha_{1},..., s_{n-1}, \alpha_{n-1}, s_{n}) \ s.t. \nonumber \\
s_{i} \stackrel{\alpha_{i}}{\rightarrow} s_{i+1} \mbox{ for all } i<n. \nonumber
\end{eqnarray}
%Clearly, not every agent is a participant in every action in the history. We acknowledge this by defining $\alpha^{A}$, or the way agent $A$ perceives actions $\alpha$. 
%\runinhead{Definition}[Appearance of actions]\\
%$\alpha^{A} =
% \begin{cases} \ \alpha & \mbox{ if } A:_{s}\alpha \\ 
% \{\beta \in i\mathcal{A}c \mbox{ s.t. } \neg A:_{s}\beta \} & \mbox{if  } \neg A:_{s}\alpha 
% \end{cases}$ 

%If $A$ is not a participant, then $\alpha$ could as well be any other action $\beta$ in which $\alpha$ does not participate.\\
%\begin{center}
%$h^{A}_{i} \ = \ (s_{0}(A), \alpha_{0}^{A}, s_{1}(A), \alpha_{1}^{A},..., s_{i-1}(A), \alpha_{i-1}^{A}, s_{i}(A))$
%\end{center}

%For any $h = (s_{0}, \alpha_{0}, s_{1}, \alpha_{1},..., s_{n-1}, \alpha_{n-1}, s_{n}),$
\begin{itemize}
\item
$(h, \alpha, t) := (s_{0}, \alpha_{0}, s_{1}, \alpha_{1},..., s_{n-1}, \alpha_{n-1}, s_{n}, \alpha, t)$ iff $s_{n} \stackrel{\alpha}{\rightarrow} t$
\item
$| h |$ denotes the size of the history, equal to the number of state-action pairs in the history, not counting the final state
\item
$last(h) = s_{n}$. We use the convention of $h\models\varphi \ $ iff $ \ last(h)\models\varphi$, read ``history $h$ satisfies statement $\varphi$ if and only if the last state in history $h$ satisfies statement $\varphi$"
\end{itemize}

We extend the notion of state indistinguishability to history indistinguishability for an agent $A$.

\runinhead{Definition}[Equivalence of histories]\\
Let $h = (s_{0}, \alpha_{0}, s_{1}, \alpha_{1},..., s_{i}, \alpha_{i},..., \alpha_{n-1}, s_{n})$ and \\
let $h' = (s'_{0}, \alpha'_{0}, s'_{1}, \alpha'_{1},..., s'_{i}, \alpha'_{i},..., \alpha'_{n-1}, s'_{n})$, then \\
\par 
$h \stackrel{A}{\sim} h' \Leftrightarrow \forall i \in \{0,1, 2, .... n\}: \ |h|=|h'| \mbox{ and } s_{i} \stackrel{A}{\sim} s'_{i} \mbox{ and } \alpha_{i} \stackrel{A}{\sim}_{s_{i}} \alpha'_{i}$

\runinhead{Definition}[History transition]\\
For two histories $h, h'$, \\
$h \stackrel{\alpha}{\rightarrow} h'$ iff $\exists  t \ s.t. \ h'=(h,\alpha,t).$  \par

\runinhead{Proposition}[Perfect Recall] This follows from the definitions above and ensures uniqueness of history transitions.\\
$h'\sim_{C}h''$, $h'=(h_{1}, \alpha, s')$, $h''=(h_{2}, \beta, s'') \ \Rightarrow \ |h_{1}|=|h_{2}|$, $h_{1}\sim_{C}h_{2}$, $\alpha\sim_{C}\beta$.

\runinhead{Proposition}Indistinguishable histories for an agent remain indistinguishable for its subagents in the last state: \par
$h\sim_{C}h', \ A<^{+}_{last(h)}C \ \Rightarrow \ h\sim_{A}h'$
\begin{proof}
\smartqed
$s_{i}(X)=s'_{i}(X)$, for all $i$, for all $X\leq^{+}C$ implies the same for $X\leq^{+}A$ since $A$ is a subagent of $C$.\\
Now, for each $\alpha_{i}\sim_{C}\alpha'_{i}$ in the histories, if $C$ is not a participant of $\alpha$ and they appear to be the same, then by definition of participation the same holds for $A$ since it is a subagent.\\
If $C:_{s_{i}}\alpha$, then $\alpha_{i}=\alpha'_{i}$. In this case, regardless of whether or not $A$ participates in $\alpha$, the two appear the same to it. 
\qed
\end{proof}

 The definition for equivalence of histories for a group of agents is similar:  
\begin{equation}
h \sim_{B_{1}, ..., B_{n}} h' \stackrel{def}{:=} h\sim_{B_{1}}h' \cap \ ...\ \cap h\sim_{B_{n}}h' \label{capsinh} 
\end{equation}

\subsection{Semantics}
The semantics of our language is embodied by a satisfaction relation $\models$ between histories and logical formulas, which is defined by the inductive clauses in Table 3. The definition is by induction on formulas.
For $A,B,B_{1}, ..., B_{n}$, distinct, $\in \mathcal{A}$:
\begin{table}
\caption{Semantics}
\begin{svgraybox}
\begin{tabular}{l l l l @{}}
\hline
$h\models \ A<^{+}B  $ & iff & $ A<^{+}_{last(h)}B$ \\
%$h \models p$ &  iff & $p \mbox{ is true at }last(h)$ & $\mbox{}$ \\ 
$h \models \neg \varphi$ &  iff  & $h \not\models \varphi$ & $\mbox{}$ \\ 
$h \models  \varphi \wedge \psi$ &  iff  & $h \models \varphi \mbox{ and } h \models \psi$ & $\mbox{}$   \\ 
$h \models K_{A}\varphi$ &  iff  &  $\forall (h'\sim_{A}h): h'\models \varphi$ & $\mbox{}$ \\ 
$h\models DK_{B_{1}, ..., B_{n}}\varphi$ &  iff  & $ \forall (h' \sim_{B_{1}, ..., B_{n}} h): \ h' \models \varphi $ & $\mbox{}$ \\
$h\models[\alpha]\varphi$ &  iff  &  $\forall  h\stackrel{\alpha}{\rightarrow}h': \ h'\models\varphi$ & $\mbox{}$ \\
\hline
\end{tabular}
\end{svgraybox}
\end{table}

%\subsection{Knowledge}
%Given the indistinguishability relations above, we define knowledge (K) of an agent $A$ and distributed knowledge (DK) of a group of agents. 
%\begin{eqnarray}
%h \models K_{A}\varphi \mbox{ iff } \forall h\stackrel{A}{\sim}h', h'\models \varphi \nonumber \\
%h\models DK_{B_{1}, ..., B_{n}}\varphi \ \mbox{ iff } \ \forall (h' \sim_{B_{1}, ..., B_{n}} h), \ h' \models \varphi \nonumber
%\end{eqnarray}
%
%Note the relationship between equivalence relations and DK for different group sizes. For $B_{1},...,B_{n},A\in\mathcal{A}$:
%\begin{eqnarray}
%h\sim_{B_{1},...,B_{n},A}h' \ \subseteq \ h\sim_{B_{1},...,B_{n}}h' \label{groupeq6} \\
%DK_{B_{1},...,B_{n}}\varphi \ \Rightarrow \ DK_{B_{1},...,B_{n},A}\varphi 
%\end{eqnarray}
%We point to a duality: \textit{Intersecting equivalence relations shared by a smaller group include those shared by an expanded group, while distributed knowledge of a smaller group forms a subset of the knowledge shared by an expanded group.}

\section{Proof System}
\label{Proof System}
We use axioms and rules of inference from propositional logic and those of DEL\footnote{These include the Necessitation and the Modus Ponens rules of inference, as well as KT45 axioms and all tautologies of propositional logic. See \cite{DELDit} for more description.}, together with those specific to our formalism. In addition, we outline reduction laws, with select proofs. In this section, $A, B, C, X, Y, B_{1}, ..., B_{n},A_{1}, ..., A_{n}$ are agents $\in\mathcal{A}$.
\begin{table}
\caption{Axioms of Knowledge}
\begin{svgraybox}
\begin{tabular}{l l l l @{}}
\hline
$\vdash DK_{A}\varphi $ & $\Leftrightarrow$ & $ K_{A}\varphi$ & G1\\
$\vdash K_{A}\varphi $ & $\Rightarrow$ & $ DK_{A,B_{1},...,B_{n}}\varphi$ & KtoDK\\
$$\\
$\vdash A<^{+}C$  &  $\Rightarrow$ & $K_{C}(A<^{+}C)$ &  KOwn \\
$\vdash A<^{+}C \ \wedge \ C<^{+}B$ & $\Rightarrow$ & $DK_{B,A_{1}, ..., A_{n}}(A<^{+}C)$ & DKOwn \\
$$\\

$B_{1}, ..., B_{n} <^{+} A \ \wedge \ DK_{B_{1}, ..., B_{n}}\varphi \ $ & $ \Rightarrow $  & $\ K_{A}\varphi $ & KfromDK \\
\hline
\end{tabular}
\end{svgraybox}
\end{table}

\begin{proof}
\smartqed
[KOwn] The right hand side of the statement is equivalent to $\forall h' (h \sim_{C} h'  \ \Rightarrow \ h'\models A<^{+}C)$. 
By the definition of equivalence, we have that $\forall i, \ s_{i}(X)=s_{i}'(X)$, for all $X\leq^{+} C$, which implies that state assignments, for all states in histories $h,h'$ will be the same for $C$ and its subagents. \\
But then $last(h)(X)=last(h')(X)$ will also hold true for $X=C$ and $X=A$ and all agents in between them, thus satisfying $h'\models A<^{+}C$. 
\qed
\end{proof}

Axioms R, Trans, and Tree reveal the loop-less tree structure of agents.
\begin{svgraybox}
\runinhead{Axiom R}
$\vdash \neg A<^{+}A $
\runinhead{Axiom Trans}
$\vdash A<^{+}B \wedge B<^{+}C \ \Rightarrow \ A<^{+}C$ 
\runinhead{Axiom Tree} 
$\vdash (X<^{+}A \wedge X<^{+}B) \ \Rightarrow \ (A<^{+}B \vee B<^{+}A) $\\
\end{svgraybox}
\begin{proof} 
\smartqed
[Axiom Tree] For $s=last(h)$, the statement is semantically equivalent to $X<^{+}_{s}A $ and $ X<^{+}_{s}B$, for some state $s$. But then by \eqref{treelong}, we guarantee that $B<^{+}_{s}A $ or $A<^{+}_{s}B$, which is semantically equivalent to the desired result.
\qed
\end{proof}

We now explore reduction laws involving the dynamic modality.
\begin{svgraybox}
\runinhead{Partial Functionality Axiom}
$ [\alpha]\neg\varphi \Leftrightarrow (<\alpha>\top \Rightarrow \neg[\alpha]\varphi)$
\end{svgraybox}
That is, the transition induced by $\alpha$, if it exists, goes to a unique next state: if $h\stackrel{\alpha}{\to}h'$ and $h\stackrel{\alpha}{\to}h''$, then $h'=h''$. This ensures uniqueness of transition.

\par
The Preservation of Facts axiom of DEL demands several versions for the different types of actions (see Table 5). 
\begin{table}
\caption{Preservation of Facts Axioms$^a$} 
\begin{svgraybox}
\begin{tabular}{l l l l @{}}
\hline
$[\alpha_{I}]\varphi$ & $\Leftrightarrow$ & $ (<\alpha_{I}>\top \Rightarrow \varphi) $ & PF1 \\
%$$\\
%$<\alpha_{II}>\top $ & $ \Rightarrow $ & $ A<E \wedge C<E$ & EXII\\
%$<\alpha_{III}>\top $ & $ \Rightarrow $ & $ A<C \wedge C<E$ & EXIII\\
%$<\alpha_{IV}>\top $ & $ \Rightarrow $ & $ A<E \wedge C<E$ & EXIV\\
%$[\alpha_{II}]A<C $ & $  $ & $ $ & AXII\\
%$[\alpha_{III}]A<E $ & $  $ & $ $ & AXIII\\
%$[\alpha_{IV}]\neg C<E $ & $  $ & $ $ & AXIV\\
%$$\\
For $X\neq A$:\\
$[\alpha_{II}]X<Y $ & $ \Leftrightarrow \ $ & $  (<\alpha_{II}>\top \Rightarrow \ X<Y)$ & PF2a\\
For $Y\neq E,C$:\\
$[\alpha_{II}]A<Y $ & $ \Leftrightarrow \ $ & $  (<\alpha_{II}>\top \Rightarrow \ A<Y)$ & PF2b\\
%$$\\
For $X\neq A$:\\
$[\alpha_{III}]X<Y $ & $ \Leftrightarrow \ $ & $  (<\alpha_{III}>\top \Rightarrow \ X<Y)$ & PF3a\\
For $Y\neq E,C$:\\
$[\alpha_{III}]A<Y $ & $ \Leftrightarrow \ $ & $  (<\alpha_{III}>\top \Rightarrow \ A<Y)$ & PF3b\\
%$$\\
$(X<C) $ & $\Rightarrow $ & $ [\alpha_{IV}](X<A) $ & PF4a\\
For $X\neq C$: \\
$\neg(X<A) \ \Rightarrow \  ( [\alpha_{IV}]X<Y $ & $ \Leftrightarrow \ $ & $ (<\alpha_{IV}>\top \Rightarrow \ X<Y)) $ & PF4b  \\
$$\\
\hline
\end{tabular} 
\end{svgraybox}
$^a$ Note that the Consequences outlined in Table 2 also belong to this category of reduction laws.
\end{table}

\begin{proof}
\smartqed
[PF4a] We unwrap the definition for Type IV action, found in \ref{transitions}, where $s=last(h)$. It follows: \\
If $X<C$ at $last(h)$, then $X\sqsubseteq S$ (occurs in process $S$). \\
Since $s'(X)=s(X),$ for all $X\neq A,C,E$, then $X$ still occurs in $S$ at $s'$.\\
Since $s'(A)$ is assigned process $S$, where $X$ occurs, then $X$ must be a subagents of $A$ at $s'$.
\qed
\end{proof}

Similarly, the Action-Knolwedge reduction laws are expanded for specificity (see Table 6).
\begin{table}
\caption{Action-Knowledge Axioms}
\begin{svgraybox}
\begin{tabular}{l l l l l @{}}
\hline
\\
For $X=A,C$: & $[\alpha_{I}]K_{X}\varphi$ & $ \Leftrightarrow$ & $ (<\alpha_{I}>\top \Rightarrow K_{X}[\alpha_{I}]\varphi)$ &  AcKn1a\\
$(A<^{+}X \vee C<^{+}X) \Rightarrow$ & $[\alpha_{I}]K_{X}\varphi$ & $  \Leftrightarrow $ & $(<\alpha_{I}>\top \Rightarrow K_{X}[\alpha_{I}]\varphi)$ & AcKn1b\\
$$\\
$$&$[\alpha_{II}]K_{C}\varphi$ & $ \Leftrightarrow $ & $(<\alpha_{II}>\top \Rightarrow DK_{A,C}[\alpha_{II}]\varphi)$ & AcKn2a\\
$C <^{+} X  \Rightarrow$ & $ [\alpha_{II}]K_{X}\varphi $ & $\Leftrightarrow$ & $ (<\alpha_{II}>\top \Rightarrow K_{X}[\alpha_{II}]\varphi)$ & AcKn2b\\
$$&$[\alpha_{II}]K_{A}\varphi $ & $\Leftrightarrow$ & $ (<\alpha_{II}>\top \Rightarrow K_{A}[\alpha_{II}]\varphi)$ & AcKn2c\\
$$\\
For $X=A,C$: & $[\alpha_{III}]K_{X}\varphi$ & $ \Leftrightarrow$ & $ (<\alpha_{III}>\top \Rightarrow K_{X}[\alpha_{III}]\varphi)$ &  AcKn3a\\
$A<^{+}X  \Rightarrow$ & $ [\alpha_{III}]K_{X}\varphi$ & $ \Leftrightarrow $ & $(<\alpha_{III}>\top \Rightarrow K_{X}[\alpha_{III}]\varphi)$ & AcKn3b\\
$$\\
$$&$[\alpha_{IV}]K_{A}\varphi $ & $\Leftrightarrow$ & $ (<\alpha_{IV}>\top \Rightarrow DK_{A,C}[\alpha_{IV}]\varphi)$ & AcKn4a\\
$A<^{+}X \Rightarrow$ & $ [\alpha_{IV}]K_{X}\varphi$ & $ \Leftrightarrow$ & $ (<\alpha_{IV}>\top \Rightarrow K_{X}[\alpha_{IV}]\varphi)$ &  AcKn4b\\
$$\\
$(X<^{+}A \wedge X<^{+}C)\Rightarrow $&$ [\alpha]K_{X}\varphi $ & $ \Leftrightarrow $ & $ \bigwedge_{\beta \in i\mathcal{A}c^{+}, \beta\sim_{X}\alpha} (<\alpha>\top \ \Rightarrow [\beta]\varphi)$ &      AcKnNP\\
\hline
\end{tabular}
\end{svgraybox}
\end{table}
\par
Note that the final rule in Table 6 is for non-participants of any action $\alpha$. All proofs are achieved by a counterfactual argument of ``chasing the diagram,'' though we omit them here.

\runinhead{Theorem}
The proof system for $\mathcal{PADEL}$ is sound.
\begin{proof}
\smartqed
In order to show soundness, all axioms in the system must be valid. For all axioms presented in gray boxes, validity was either proved in the text or it follows from the semantic definitions.
\qed 
\end{proof}

\runinhead{Theorem}[Model-checking]
The model-checking problem for $\mathcal{PADEL}$ is decidable on finite models.
\begin{proof}
\smartqed
Given a model $M$ with a countable set of histories $h$ and formula $\varphi$, the axioms and rules of inference are sufficient to decide whether or not $\varphi$ is satisfiable at $M,h$, since we have provided axioms for all syntactic combinations of terms $\varphi$ can have.
\qed
\end{proof}

\runinhead{Corrollaries}
The following are semantically valid consequences of axioms and rules of inference:
\begin{svgraybox}
\begin{itemize}
\item
$\vdash A<B \ \Rightarrow \ A<^{+}B $ 
\item
$\vdash X<A \ \Rightarrow \ \neg{X<B}$ 
\item
$\vdash A<^{+}C \ \wedge \ C<^{+}B_{1}, ..., B_{n} \ \Rightarrow \ DK_{B_{1}, ..., B_{n}}(A<^{+}C)$ 
\item
$\vdash A<^{+}C \ \wedge \ C<^{+}B \ \Rightarrow \ K_{B}(A<^{+}C) $
%\item
%$B<^{+}A \ \wedge \ K_{B}\varphi \ \Rightarrow \ K_{A}\varphi$ 
\item
$\vdash B_{1}, ..., B_{n} <^{+} A \ \wedge \ DK_{B_{1}, ..., B_{n}, A}\varphi \ \Rightarrow \ K_{A}\varphi $ 
%\item
%$\vdash [\alpha_{II}](\neg A<E)$ 
%\item
%$(X<^{+}A) \ \Rightarrow \ [\alpha_{II}](X<^{+}C)$ 
%\item
% $\vdash [\alpha_{III}]\neg(A<C)$ 
\item
$(X<^{+}A) \ \Rightarrow \  [\alpha_{III}]\neg(X<^{+}C)$
%\item
%$(X<C) \ \Rightarrow \ [\alpha_{IV}] \neg(X<C)$
\end{itemize}
\end{svgraybox}

%\runinhead{Open Questions}
%\begin{itemize}
%\item
%It remains to investigate whether the system is decidable (that is, whether, given a formula $\varphi$, there exists a model $M$ and a history $h$ s.t. $M,h\models\varphi$).
%\item
%It remains also to determine completeness of the system. We predict that the system may need more axioms added before completeness can be shown.
%\end{itemize}

\section{Conclusion}

We have thus developed a sound, decidable language $\mathcal{PADEL}$ based on a nested tree structure of a finite number of agents, which are defined by concurrent processes, subagents and capabilities. Furthermore, we developed the notion of knowledge and distributed knowledge for agents based on
\begin{enumerate}
\item
the currect \emph{state} of an agent, which captures its current $one-step$ subagents and its current capabilities for future interactions
\item 
the current \emph{state} of all of its $iterative$ subagents. This encodes a principle of monotonicity of information: all information carried by a subagent is available to any of its superagents
\item
the memory of an agent, encoded in a history that each agent perceives differently. Following the premises of DEL, information is never lost and contradictory knowledge is never acquired.
\end{enumerate}
The presented axiomatization allows one to reason about knowledge and change in knowledge of agents executing actions, as well as their subagents and superagents. Further applications to biological systems remain to be explored, in particular seeking to define ``knowledge,'' as described by indistinguishabilities, for a given biological unit. It also remains to investigate whether the system is complete.

\begin{acknowledgement}
The first author's contribution to this paper is part of her Master of Science thesis, which was supported by the Rhodes Trust. Her presentation of $\mathcal{PADEL}$ at ECAL11, which brough about this paper, was made possible by Amgen. For inquiries and proofs not presented in this paper, please contact her at ayermakova@gmail.com
\end{acknowledgement}

\end{document}